\documentclass[twocolumn,prb,showpacs]{revtex4} 
\usepackage{mathrsfs} 
\usepackage{amssymb} 
\usepackage{graphicx}
\usepackage{subfigure}
\usepackage{dcolumn}
\usepackage{bm}
\begin{document}

\title{Electron-Phonon Interactions for Optical Phonon Modes in Few-Layer Graphene}
\author{Jia-An Yan, W. Y. Ruan, and M. Y. Chou}
\affiliation{School of Physics, Georgia Institute of Technology,
Atlanta, Georgia 30332 USA}
\date{\today}
\begin{abstract}
We present a first-principles study of the electron-phonon (e-ph)
interactions and their contributions to the linewidths for the
optical phonon modes at $\Gamma$ and K in one to three-layer
graphene. It is found that due to the interlayer coupling and the
stacking geometry, the high-frequency optical phonon modes in
few-layer graphene couple with different valence and conduction
bands, giving rise to different e-ph interaction strengths for these
modes. Some of the multilayer optical modes derived from the
$\Gamma$-$E_{2g}$ mode of monolayer graphene exhibit slightly higher
frequencies and much reduced linewidths. In addition, the linewidths
of K-$A'_1$ related modes in multilayers depend on the stacking
pattern and decrease with increasing layer numbers.

\end{abstract}

\maketitle

The possibilities of developing carbon-based nanostructures for
electronics applications have stimulated recent interest in graphene
and its derivatives \cite{Novoselov2004,Berger2004,Zhang2005}. One
of the focus areas is to understand the scattering processes of
electrons. In carbon nanotubes and graphite, the high currents or
optical excitations have been shown to induce a significant
overpopulation of the optical phonon modes of $E_{2g}$ at $\Gamma$
($\Gamma$-$E_{2g}$) and $A_1'$ at K (K-$A_1'$)
\cite{Yao2000,Lazzeri2005,Kampfrath2005}. Since these phonon modes
exhibit a strong electron-phonon (e-ph) interactions, overpopulation
of them leads to a dramatic reduction of the ballistic conductance
of carbon nanotubes at high bias potentials \cite{Yao2000}, and
consequently interconnect performance deteriorates
\cite{Bonini2007}. Understanding the phonon decays
\cite{Grimvall1981} from a microscopic point of view, in particular,
based on the e-ph interaction, is thus a key step to improve the
transport properties of these materials and to control device
performance. Furthermore, the e-ph interaction also plays a
significant role in many phenomena such as the quasiparticle
dynamics and anomalies in photoemission spectra
\cite{Bostwick2007ssc, Gonzalez2008, Calandra2007, Zhou2008,
Hwang2008}, Raman scattering \cite{Ferrari2007}, and
superconductivity \cite{Giustino2007}.

Experimentally, the linewidths of the zone-center phonon modes
obtained from Raman or infared (IR) measurements contain significant
contributions from the e-ph interaction \cite{Menendez1984}. In
graphene, the phonon linewidth of the $\Gamma$-$E_{2g}$ mode is
estimated to be about 13 cm$^{-1}$ based on the Raman spectra
\cite{Yan2007}. In graphite, the graphene $E_{2g}$ mode splits into
two branches: the Raman-active $E_{2g}$ and IR-active $E_{1u}$
modes. The linewidth of the Raman-active mode (11.5 cm$^{-1}$
\cite{Lazzeri2006}) is almost the same as that of graphene, while IR
measurements show that the linewidth of the $E_{1u}$ mode is
surprisingly much smaller \cite{Nemanich1977}.

\begin{figure}[tbp]
\centering
   \includegraphics[width=8.0cm]{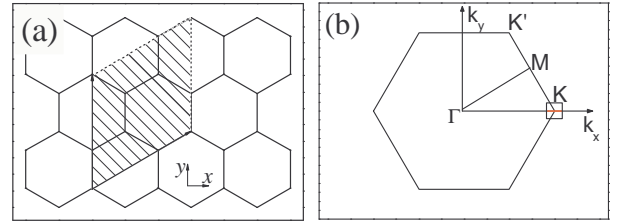}
 \caption{(a) Supercell (shaded area) used in the frozen-phonon calculation for
 the K phonon modes in few-layer graphene. (b) First Brillouin zone of the primitive unit cell
 of graphene. The square area for the dense $\bf k$-grid sampling is indicated. \label{fig:model}}
\end{figure}

Few-layer graphene (FLG) presents an interesting system because of
the possibility to tune its electronic properties \cite{Ohta2006}.
In the epitaxially grown graphene, FLG is often produced as a main
product \cite{Berger2004}. Depending on the layer number and
stacking geometry \cite{Latil2006}, the linear band dispersions in
monolayer graphene evolve into several bands due to the interlayer
coupling in FLG \cite{Latil2006,Ohta2007}. Similarly, the $E_{2g}$
mode at $\Gamma$ and $A_1'$ mode at K in single layer also split
into several branches \cite{me2007}. It is expected that the e-ph
interaction will be significantly modified as the number of layers
and stacking geometry are varied.

In this work, we performed first-principles calculations of the e-ph
interactions in one-, two- (AB stacking), and three-layer (ABA and
ABC stackings) graphene. The $\Gamma$-$E_{2g}$ and K-$A_1'$ modes in
graphene, which are respectively coupled to intra- and inter-valley
electronic scatterings, have much stronger e-ph interactions than
other modes \cite{Tse2008}. Here we focus on these two modes and
their derivatives in FLGs in order to investigate the stacking
effect. We found that while the weak interlayer interaction gives
rise to only small splittings of these phonon modes, the resulting
e-ph interactions for some of them are considerably suppressed
because of symmetry constraints.

\begin{figure*}[tbp]
\centering
   \includegraphics[width=15.0cm]{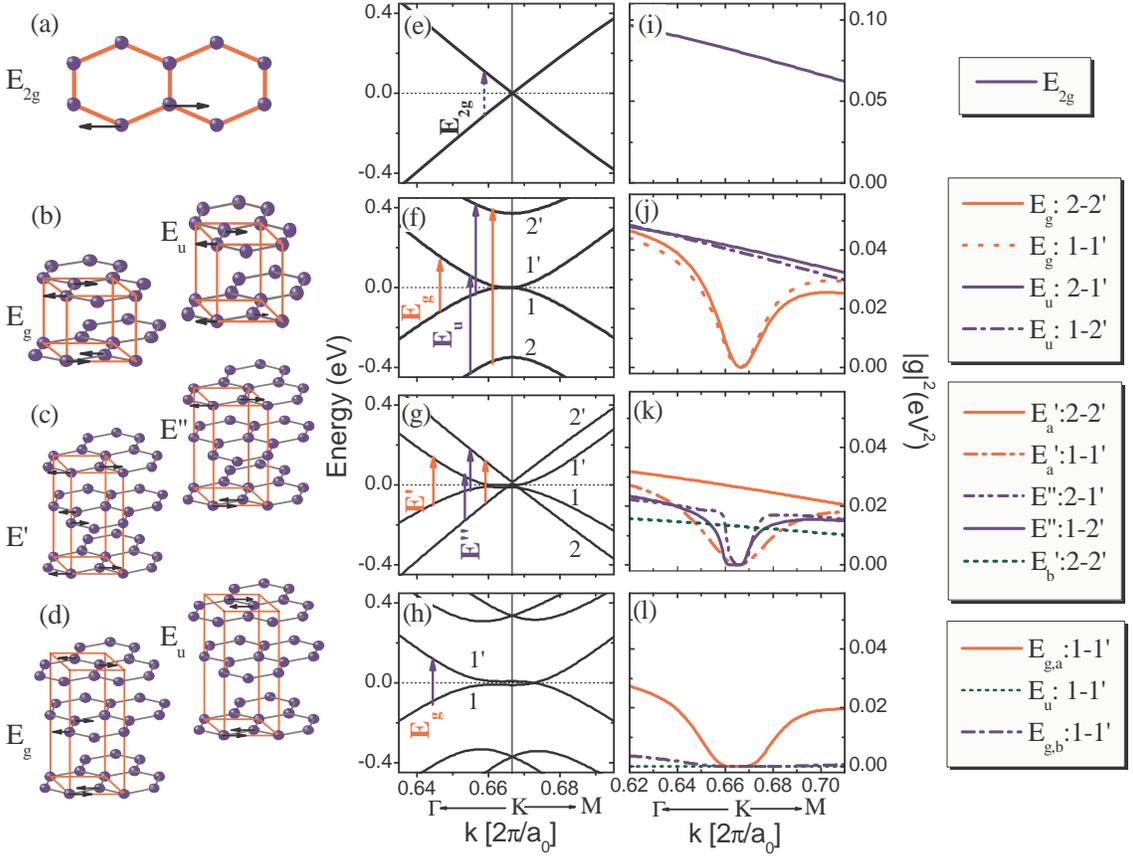}
 \caption{(Color online) (a)-(d) Optical modes at $\Gamma$, (e)-(h) band dispersions near the Fermi level, and (i)-(l) the square of the e-ph interaction
 strength $|g|^2$ of the optical modes at $\Gamma$ for monolayer, AB bilayer, ABA and ABC trilayer graphene, respectively. Symmetry-allowed transitions from valence bands to conduction bands with nonzero strength are shown.
 \label{fig:fig4}}
\end{figure*}

The e-ph matrix element $g_{(\mathbf{k+q})j',\mathbf{k}j}^{\nu}$ is
defined as
\begin{eqnarray}
g_{(\mathbf{k+q})j',\mathbf{k}j}^{\nu}=\sqrt{\frac{\hbar}{2M\omega_{\mathbf{q}}^{\nu}}}\langle
\mathbf{k+q},j'|\frac{\delta V_{scf}}{\delta
u_{\mathbf{q}}^{\nu}}|\mathbf{k},j\rangle,
\end{eqnarray}
where $\delta V_{scf}\equiv
V_{scf}(u_{\mathbf{q}}^{\nu})-V_{scf}(0)$ is the variation of the
self-consistent potential field due to the perturbation of a phonon
with wave vector $\mathbf{q}$ and branch index $\nu$.
$|\mathbf{k},j\rangle$ is the electronic Bloch state.  We further
define the e-ph coupling strength between phonon mode
$\mathbf{q}\nu$ and Bloch states $|\mathbf{k+q},j'\rangle$ and
$|\mathbf{k},j\rangle$ as
$g_{j'j}^2(\mathbf{k})=|g_{(\mathbf{k+q})j',\mathbf{k}j}^{\nu}|^2$.

The electronic states and the self-consistent field were computed
using the first-principles codes Quantum ESPRESSO \cite{Baroni2001,
pwscf} and the perturbation of a phonon mode was handled with the
frozen-phonon approach. Fig.~\ref{fig:model}(a) shows the supercell
used in the frozen-phonon calculation for the K phonons. The
electronic structure was calculated with the local density
approximation (LDA) within density-functional theory, and the
core-valence interaction was modeled by norm-conserving
pseudopotentials \cite{Troullier1991}. Wave functions of the valence
electrons were expanded in plane waves with a kinetic energy cutoff
of 70 Ry. The phonon frequencies and associated eigenvectors were
computed using the density-functional perturbation theory (DFPT)
\cite{Baroni2001}, details of which have been presented in our
previous work \cite{me2007}. A vacuum region of 10 \AA~ was
introduced in our supercell to eliminate the artificial interaction
between neighboring supercells along the $z$ direction. The relaxed
C-C bond length is 1.42 \AA~  and the interlayer distance is 3.32
\AA~ for all FLGs considered in this paper. Variations of the
potential fields $\delta V_{scf}$ were calculated through
self-consistent calculations to find the potential field for both
perturbed and unperturbed systems. The following calculations were
carried out on a dense 100$\times$100 $k$-grid within a small square
area enclosing point K in reciprocal space, as indicated in
Fig.~\ref{fig:model}(b) in order to obtain the electronic wave
functions at $\mathbf{k}$ and $\mathbf{k+q}$ near the Fermi level.
This dense $k$-sampling was found necessary for a quantitative
description of the scattering process near the Fermi level in one-
and few-layer graphene. Finally, the e-ph interaction matrix
elements were computed using Eq.~(1).

In order to check the accuracy of our calculations, we first
calculate the e-ph matrix elements over the Fermi surface and
compare them with previously published results for monolayer
graphene. Due to the electronic degeneracy at K, the averaged e-ph
matrix elements for all possible pairs are $\langle
g_{\Gamma}^2\rangle_F$=$\sum_{i,j}
|g_{(\mathbf{K})i,\mathbf{K}j}|^2/4$=0.0401 eV$^2$ for the
$\Gamma$-$E_{2g}$ mode, and $\langle
g_{\mathbf{K}}^2\rangle_F$=$\sum_{i,j}
|g_{(2\mathbf{K})i,\mathbf{K}j}|^2/4$=0.0986 eV$^2$ for the K-$A_1'$
mode, respectively. These results are in excellent agreement with
those in previous DFPT calculations (0.0405 and 0.0994 eV$^2$,
respectively) \cite{Piscanec2004}.

The phonon linewidth $\gamma$ due to the e-ph coupling is defined as
\cite{Allen1972}:
\begin{eqnarray}
\gamma^{\mathbf{q}\nu}
&=&\frac{4\pi}{N_k}\sum_{\mathbf{k}jj'}|g_{(\mathbf{k+q})j',\mathbf{k}j}^{\nu}|^2[f_{\mathbf{k}j}-f_{(\mathbf{k+q})j'}]\nonumber\\
&&\times\delta[\varepsilon_{\mathbf{k}j}-\varepsilon_{(\mathbf{k+q})j'}+\hbar\omega_{\mathbf{q}}^{\nu}]\label{eq:gamma},
\end{eqnarray}
with $f_{\mathbf{k}j}$ being the Fermi-Dirac occupation function for
Bloch state $|\mathbf{k},j\rangle$. We used a broadening parameter
of 0.01 eV for the $\delta$-function in Eq.~(\ref{eq:gamma}), and
$k_BT$=2.5 meV for the Fermi-Dirac distribution.

\begin{figure}[tbp]
\centering
   \includegraphics[width=8.0cm]{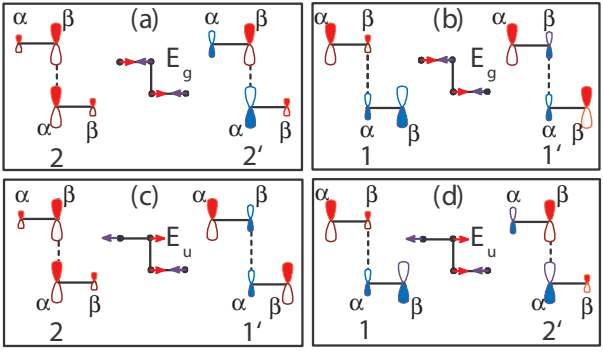}
 \caption{(Color online) Schematic diagrams of electronic orbitals and optical phonon modes at $\Gamma$
 that give rise to non-zero e-ph matrix elements in bilayer graphene. The tight-binding
 orbitals of each band close to K are shown. Red (blue) is
 positive (negative) amplitude. \label{fig:fig5}}
\end{figure}

Table~\ref{tab:freq} lists the calculated phonon linewidths due to
the e-ph coupling. For the monolayer, the linewidth $\gamma$ of the
degenerate $\Gamma$-$E_{2g}$ modes is 11.2 cm$^{-1}$, in reasonable
agreement with experimental observation of 15.0 cm$^{-1}$
\cite{Yan2007} and previously published result of 11.5 cm$^{-1}$
\cite{Lazzeri2006}. A value of 20.4 cm$^{-1}$ is obtained for the
highest optical K-$A_1'$ mode. In FLGs, the phonon modes split into
branches with different symmetries. Consequently, their linewidths
can be quite different from one another. In the AB bilayer, for
example, the original highest $\Gamma$ phonon splits into the Raman
active $E_g$ and IR-active $E_u$ modes. While the linewidth of the
$E_g$ mode (8.6 cm$^{-1}$) is only slightly smaller than that of the
monolayer (11.2 cm$^{-1}$), the linewidth of the $E_u$ mode is two
orders of magnitude smaller. The calculated linewidth of the $E_g$
mode is consistent with recent experimental observation of 13.5
cm$^{-1}$ for the $\Gamma$ optical phonon mode in bilayer graphene
\cite{Yan2008}. Similar results for the monolayer and bilayer
graphene were also reported by Park \emph{et al.} \cite{Park2008}.
For the ABA and ABC trilayer, the results in Table~\ref{tab:freq}
show different behavior for different modes. In particular, for the
ABC trilayer two out of three optical modes at $\Gamma$ show reduced
electron-phonon interaction. Most importantly, the linewidths of all
K phonons in Table~\ref{tab:freq} drop significantly as the layer
number increases from 1 to 3. This indicates that the interlayer
interaction can effectively suppress vallley-spin decoherence via
e-ph scatterings, making the FLGs more attractive as valleytronic
materials \cite{Rycerz2007,Akhmerov2007}.


\begin{table*}[tbp]
 \caption{Calculated phonon linewidth $\gamma$ (in cm$^{-1}$) for the high-frequency optical phonon modes at $\Gamma$ and K in monolayer,
bilayer, and trilayer graphene. The mode symmetries S and the
frequencies $\omega$ (in cm$^{-1}$) are also listed for
completeness.} \label{tab:freq}
\begin{ruledtabular}
\begin{tabular}{cllllllllllll}
            & \multicolumn{3}{c}{Monolayer}  & \multicolumn{3}{c}{AB} & \multicolumn{3}{c}{ABA} & \multicolumn{3}{c}{ABC}\\
            \cline{2-4} \cline{5-7} \cline{8-10} \cline{11-13}
            & S & $\omega$ & $\gamma$ & S & $\omega$ &
            $\gamma$ & S & $\omega$ & $\gamma$ & S & $\omega$ & $\gamma$\\
            \hline
$\Gamma$    &$E_{2g}$ & 1586    & 11.2 & $E_g$ & 1587 & 8.6  & $E_a'$  & 1586 & 9.7 & $E_{g,a}$ & 1586 & 7.2  \\
            &                   &      &      & $E_u$ & 1592 & 0.1  & $E''$ & 1588 & 11.0& $E_u$ & 1589 & 0.0  \\
            &                   &      &      &                 &      &      & $E_b'$  & 1593 & 2.8 & $E_{g,b}$ & 1594 & 0.3  \\
            &                   &      &      &                 &      &      &        &      &     &       &      &      \\
K           & $A_1'$  & 1306 & 20.4 & $E$ & 1318 & 9.0  & $E_1'$,$E_1''$ & 1316 & 8.4 & $E$ & 1318 & 2.8  \\
            &                   &      &      &                 &      &      & $E_2'$  & 1324 & 3.6 & $A_1$ & 1325 & 2.2  \\
\end{tabular}
\end{ruledtabular}
\end{table*}

For FLGs there are a few valence and conduction $\pi$ bands near the
Fermi energy level. The symmetry allowed interband transitions by
the absorption of a phonon are indicated in the middle column of
Fig.~\ref{fig:fig4}. To examine the contributions of different
electronic states to the e-ph coupling strengths in FLGs, we present
in the right column of Fig.~\ref{fig:fig4} the absolute value
squared of the e-ph coupling matrix elements for optical phonons at
$\Gamma$ as a function of electronic crystal momentum $\bf k$ along
the symmetry line $\Gamma$-K for all symmetry allowed transitions.
Some of these matrix elements decrease monotonically with increasing
$k_x$, as in the case of monolayer; some of them show a minimum at
K. These features are closely related to the symmetry of the
electronic states near K in the presence of interlayer interactions,
and can be further quantitatively understood using a tight-binding
model. Following the procedures in Refs.~\onlinecite{Woods2000} and
~\onlinecite{Mahan2003}, we obtain the e-ph matrix element for the
in-plane optical phonon mode $\bf q\nu$ in $L$-layer graphene as:
\begin{widetext}
\begin{eqnarray}
g_{(\mathbf{k+q})j',\mathbf{k}j}^{\nu} &=&
g_0\sum_{l=1}^{L}\{[\vec{t}(\mathbf{k})\cdot
\vec{\epsilon}_{l\alpha}^\nu(\mathbf{q})-\vec{t}(\mathbf{k+q})\cdot\vec{\epsilon}_{l\beta}^\nu(\mathbf{q})]u_{l\alpha,j'}^{*}(\mathbf{k+q})u_{l\beta,j}(\mathbf{k})
+\\
& &[\vec{t}(-\mathbf{k-q})\cdot
\vec{\epsilon}_{l\alpha}^\nu(\mathbf{q})-\vec{t}(-\mathbf{k})\cdot\vec{\epsilon}_{l\beta}^\nu(\mathbf{q})]u_{l\beta,j'}^{*}(\mathbf{k+q})u_{l\alpha,j}(\mathbf{k})\}\nonumber\\
&\equiv &g_0 U_{j'}^\dagger(\mathbf{k+q})
\Phi^\nu(\mathbf{k},\mathbf{q}) U_j(\mathbf{k})\label{eq:gtb2},
\end{eqnarray}
\end{widetext}
with $U_j^\dagger(\mathbf{k})=[u_{1\alpha,j}^*(\mathbf{k}),
u_{1\beta,j}^*(\mathbf{k}),u_{2\alpha,j}^*(\mathbf{k}),
u_{2\beta,j}^*(\mathbf{k}),...]$ being the tight-binding amplitudes
of band $j$ for each site in the unit cell.
$\vec{\epsilon}_{l\alpha}$ and $\vec{\epsilon}_{l\beta}$ are the
vibrational eigenvectors for the two atoms in layer $l$.
$\vec{t}(\mathbf{k})=\sum_{i=1}^3 \hat{\delta_i} e^{i\mathbf{k}\cdot
\mathbf{R}_i}$, where $\hat{\delta}_i$'s ($i$=1-3) are the unit
vectors connecting atom $\alpha$ in layer $l$ to its three
nearest-neighbors (NNs), and $\mathbf{R}_i$ are the lattice vectors
of the unit cells in which the three NNs are located. The constant
$g_0=J\Omega^{1/2}/(\omega_{\mathbf{q}}^\nu\sqrt{M})$ depends on the
mode frequency $\omega_{\mathbf{q}}^\nu$, the e-ph interaction
parameter $J$, the area of the unit cell $\Omega$, and the carbon
atomic mass $M$. The coupling matrix
$\Phi^\nu(\mathbf{k},\mathbf{q})$ has the form:
\begin{equation}
\Phi^\nu(\mathbf{k},\mathbf{q})=\left(
     \begin{array}{ccccc}
       \Phi_1    &         &     &    \\
                 & \Phi_2  &        &     \\
                 &         & \ddots &  \\
                 &         &        & \Phi_L \\
     \end{array}
   \right),
\end{equation}
with
\begin{widetext}
\begin{equation}
\Phi_l=\left(
     \begin{array}{cc}
       0    &   \vec{t}(\mathbf{k})\cdot
\vec{\epsilon}_{l\alpha}^\nu(\mathbf{q})-\vec{t}(\mathbf{k+q})\cdot\vec{\epsilon}_{l\beta}^\nu(\mathbf{q})         \\
     \vec{t}^*(\mathbf{k+q})\cdot
\vec{\epsilon}_{l\alpha}^\nu(\mathbf{q})-\vec{t}^*(\mathbf{k})\cdot\vec{\epsilon}_{l\beta}^\nu(\mathbf{q})       &  0  \\
     \end{array}
   \right).
\end{equation}
\end{widetext}

It is clear that the e-ph interaction strength depends on the
orbital characteristics of the initial and final electronic states
as well as the displacement pattern of the phonon mode. Note that
$\Phi_l$ is an off-diagonal matrix, coupling the electronic
component of one sublattice to that of the other sublattice in each
layer through phonon displacements in the same layer. The coupling
effects in different layers will be summed up in either a
constructive or a destructive way, depending on the relative phase
of the phonon displacements in different layers and the relative
phase of the tight-binding amplitudes in the initial and final
electronic states involved.

\begin{figure}[tbp]
\centering
   \includegraphics[width=8.0cm]{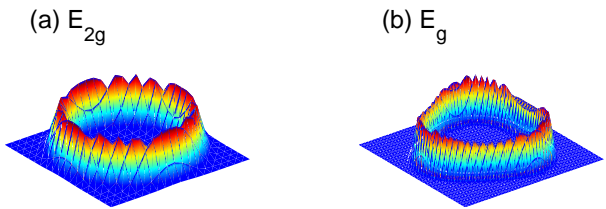}
 \caption{(Color online) Electronic momentum-resolved contributions of all scattering processes to the phonon linewidth for
 the doubly degenerate (a) $E_{2g}$ mode in monolayer graphene and (b) $E_g$ mode in bilayer graphene. Note that only -0.009$\leq$$k_x$,$k_y$$\leq$0.009 for monolayer and -0.021$\leq$$k_x$,$k_y$$\leq$0.021 for bilayer are shown. $k_x$ and $k_y$ are in $2\pi/a_0.$\label{fig:fig2}}
\end{figure}

As an example, Figure~\ref{fig:fig5} schematically shows the symmetry-allowed
transitions through the $E_g$ and $E_u$ modes at $\Gamma$ between
valence and conduction bands in bilayer graphene. The four relevant
energy bands shown in Fig.~\ref{fig:fig4}(f) are indexed as 2, 1, $1'$, and
$2'$ from lower valence bands to higher conduction bands. For the $E_g$ mode,
the in-plane vibrations of $\alpha$ and $\beta$ atoms in neighboring planes are
in phase.  Taking into account of the symmetry of electronic wave function in each layer,
allowed transitions turn out to be $1 \rightarrow 1'$ and $2 \rightarrow 2'$.
As $\bf k$ approaches K, the four bands become parabolic and the wave-function
amplitudes tend to be localized on one sublattice of each layer. Since the
transition matrix elements for 1$\rightarrow1'$ and 2$\rightarrow
2'$ contain product of the electronic wave functions on both sublattices,
small wave function amplitudes give rise to vanishingly small matrix elements near K,
as shown in Fig.~\ref{fig:fig4}(j).
On the other hand, the $E_u$ mode involves in-plane vibrations of $\alpha$ and
$\beta$ atoms in neighboring planes that are out of phase. The symmetry
allowed transitions are $2 \rightarrow 1'$ and $1 \rightarrow 2'$. Even though
the wave function amplitude at one sublattice may become small when
$\bf k$ approaches K, the products with wave function amplitudes at the other sublattice
are still finite. Therefore, the matrix elements exhibit similar linear behavior
as in monolayer graphene.

In collecting the contributions from different $\bf k$ states to
calculate the phonon linewidth in Eq.~(\ref{eq:gamma}), energy
conservation is controlled by the delta function. For the $E_g$
mode, this condition can be satisfied in the transition $1
\rightarrow 1'$ but not in the transition $2 \rightarrow 2'$.
Nevertheless, the phonon linewidth listed in Table I is about 3/4 of
the value for monolayer graphene. Figure~\ref{fig:fig2} presents the
$\bf k$-resolved linewidth contributions for the doubly degenerate
$E_{2g}$ mode in monolayer graphene and the $E_g$ mode in bilayer
graphene. Clearly, the contributions are almost isotropic in $\bf k$
space for the $E_{2g}$ mode in monolayer graphene. The shell shape
is a result of energy conservation for the scattering events. Due to
the trigonal symmetry of the constant-energy surface in bilayer
graphene, the $\bf k$-dependent contributions show a three-fold
symmetry for the $E_g$ mode. In contrast, the symmetry allowed
transitions of $2 \rightarrow 1'$ and $1 \rightarrow 2'$ for the
$E_u$ mode cannot satisfy the required energy conservation because
the energy separation of the initial and final states is too large.
This leads to a vanishingly small probability for the $E_u$ mode to
decay through the e-ph interaction. These results are a reminiscence
of the properties of $\Gamma$-$E_{2g}$ and $\Gamma$-$E_{1u}$ phonons
in bulk graphite regarding their e-ph interactions
\cite{Bonini2007}.

\begin{figure}[tbp]
\centering
   \includegraphics[width=8.5cm]{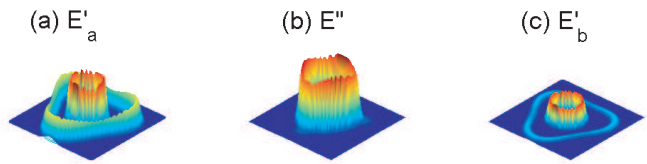}
 \caption{(Color online) Electronic momentum-resolved contributions of all scattering processes to the phonon linewidth for the (a) $E_a'$, (b) $E''$, and (c)
  $E_b'$ phonon modes at $\Gamma$ in ABA trilayer graphene. The $\bf k$ area shown in the figure is indicated in Fig.~\ref{fig:model}(b). The origin corresponds to K. \label{fig:fig3}}
\end{figure}

For the ABA trilayer, the four energy bands crossing or touching the
Fermi energy are again indexed as 2, 1, $1'$ and $2'$ in Fig.~\ref{fig:fig4}.
When a $\Gamma$-$E'_a$ phonon is absorbed, both symmetry and energy allowed
transitions are 2$\rightarrow$$2'$ and 1$\rightarrow$$1'$, which
manifest themselves as the inner and outer rings respectively in
Fig.~\ref{fig:fig3}(a). When a $\Gamma$-$E''$ phonon is absorbed,
the symmetry and energy allowed transitions are 2$\rightarrow1'$ and
1$\rightarrow2'$. Incidentally, the energy differences between bands
2 and $1'$ and between bands 1 and $2'$ are equal for any $\bf k$.
Therefore, the rings associated with these two different transitions
overlap in Fig.~\ref{fig:fig3}(b). The electronic states involved span an energy range of
$\hbar \omega \sim$ 0.2 eV about the Fermi level. A $\Gamma$-$E'_b$ phonon obeys
the same selection rule as a $\Gamma$-$E'_a$ phonon since both bear
the same group representation. The e-ph coupling strength of the
former is, however, much weaker than that of the latter.

For the ABC trilayer, the relevant energy bands are indexed as 1 and
$1'$ in Fig.~\ref{fig:fig4}. The three $\Gamma$ phonons derived from the $E_{2g}$ mode in
monolayer graphene are respectively denoted by $\Gamma$-$E_{g,a}$,
$\Gamma$-$E_u$, and $\Gamma$-$E_{g,b}$ in the order of increasing
frequencies. Interestingly, only the $\Gamma$-$E_{g,a}$ phonons show an appreciable
e-ph coupling corresponding to the transition of $1 \rightarrow 1'$.

In summary, we have studied the phonon linewidths of the
high-frequency optical phonon modes in few-layer graphene. We found
that there is a strong suppression of the e-ph interaction for these
modes resulting from stacking patterns. The split optical phonon modes in
few-layer graphene are shown to only couple with the electronic bands of
specific orbital symmetry and exhibit various e-ph interaction
strengths. These features are well illustrated using a tight-binding
model.

We acknowledge helpful discussions with M. Wierzbowska, S. Piscanec.
J.A.Y thanks F. Giustino and C.-H. Park for a critical reading of
the manuscript. This work is supported by the Department of Energy
(Grant No. DE-FG02-97ER45632) and by the National Science Foundation
(Grants No. DMR-08-20382). The computation used resources of the
National Energy Research Scientific Computing Center (NERSC), which
is supported by the U.S. Department of Energy (Grant No.
DE-AC03-76SF00098), and San Diego Supercomputer Center (SDSC) at
UCSD.


\begin{thebibliography}{99}

\bibitem{Novoselov2004} K. S. Novoselov, A. K. Geim, S. V. Morozov, D. Jiang, Y. Zhang, S. V. Dubonos, I. V. Grigorieva, A. A. Firsov, Science \textbf{306}, 666 (2004).

\bibitem{Berger2004} C. Berger, Z. M. Song, T. B. Li, X. B. Li, A. Y. Ogbazghi, R. Feng, Z. T. Dai, A. N. Marchenkov, E. H. Conrad, P. N. First, and W. A. de Heer, J. Phys. Chem.
B \textbf{108}, 19912 (2004).

\bibitem{Zhang2005} Y. Zhang, J. P. Small, W. V. Pontius, and P. Kim, Appl. Phys. Lett. \textbf{86}, 073104 (2005).

\bibitem{Yao2000} Z. Yao, C. L. Kane, and C. Dekker, Phys. Rev. Lett. \textbf{84}, 2941 (2000).


\bibitem{Lazzeri2005} M. Lazzeri, S. Piscanec, F. Mauri, A. C. Ferrari, and J. Robertson, Phys. Rev. Lett. \textbf{95}, 236802 (2005).



\bibitem{Kampfrath2005} T. Kampfrath, L. Perfetti, F. Schapper, C. Frischkorn, and M. Wolf, Phys. Rev. Lett. \textbf{95}, 187403 (2005).

\bibitem{Bonini2007} N. Bonini, M. Lazzeri, N. Marzari, and F. Mauri, Phys. Rev. Lett. \textbf{99}, 176802 (2007).

\bibitem{Grimvall1981} G. Grimvall, \emph{The Electron-Phonon Interaction in Metals}
(North-Holland, Amsterdam, 1981).




\bibitem{Bostwick2007ssc} A. Bostwick, T. Ohta, J. L. McChesney, T. Seyller, K. Horn, and E. Rotenberg, Solid State Comm. \textbf{143}, 63 (2007).



\bibitem{Gonzalez2008} J. Gonz\'{a}lez and E. Perfetto, Phys. Rev. Lett. \textbf{101}, 176802
(2008)

\bibitem{Calandra2007} M. Calandra and F. Mauri, Phys. Rev. B \textbf{76}, 205411
(2007).

\bibitem{Zhou2008} S. Y. Zhou, D. A. Siegel, A. V. Fedorov, A. Lanzara, Phys. Rev. Lett. \textbf{101}, 086402 (2008).

\bibitem{Hwang2008} E. H. Hwang and S. Das Sarma, Phys. Rev. B \textbf{77}, 081412(R) (2008).

\bibitem{Ferrari2007} A. C. Ferrari, Solid State Comm. \textbf{143}, 47 (2007).

\bibitem{Giustino2007} F. Giustino, J. R. Yates, I. Souza, M. L. Cohen, and S. G. Louie, Phys. Rev. Lett. 98, 047005 (2007).


\bibitem{Menendez1984} J. Menendez and M. Cardona, Phys. Rev. B \textbf{29}, 2051 (1984).

\bibitem{Yan2007} J. Yan, Y. Zhang, P. Kim, and A. Pinczuk, Phys. Rev. Lett. \textbf{98}, 166802 (2007).

\bibitem{Lazzeri2006} M. Lazzeri, S. Piscanec, F. Mauri, A. C. Ferrari, and J.
Robertson, Phys. Rev. B \textbf{73}, 155426 (2006).

\bibitem{Nemanich1977} R. J. Nemanich, G. Lucovsky, S. A. Solin, Solid Stat. Comm. \textbf{23}, 117 (1977).


\bibitem{Ohta2006} T. Ohta, B. Bostwick, T. Seyller, K. Horn, and E. Rotenberg, Science \textbf{313}, 951 (2006).


\bibitem{Latil2006} S. Latil, and L. Henrard, Phys. Rev. Lett. \textbf{97}, 036803
(2006).


\bibitem{Ohta2007} T. Ohta, A. Bostwick, J. L. McChesney, T. Seyller, K. Horn, and E. Rotenberg, Phys. Rev. Lett. 98, 206802
(2007).


\bibitem{me2007} J. A. Yan, W. Y. Ruan, and M. Y. Chou, Phys. Rev. B \textbf{77}, 125401 (2008)


\bibitem{Tse2008}  We note there are some further discussions of Kohn anomaly in one and bilayer graphene recently.
See, e.g., W.-K. Tse, BenYu-Kuang Hu, and S. Das Sarma, Phys. Rev.
Lett. \textbf{101}, 066401 (2008). E. H. Hwang, and S. DasSarma,
Phys. Rev. Lett. \textbf{101}, 156802 (2008).




\bibitem{Baroni2001} S. Baroni, S. de Gironcoli, and A. Dal Corso, Rev. Mod.
Phys. \textbf{73}, 515 (2001).

\bibitem{pwscf} S. Baroni, A. Dal Corso, S. de Gironcoli, and P. Giannozzi, http://www.pwscf.org.

\bibitem{Troullier1991}  N. Troullier and J. L. Martins, Phys. Rev. B \textbf{43}, 1993 (1991).

\bibitem{Piscanec2004} S. Piscanec, M. Lazzeri, F. Mauri, A. C. Ferrari, and J.
Robertson, Phys. Rev. Lett. \textbf{93}, 185503 (2004).

\bibitem{Allen1972} P. B. Allen, Phys. Rev. B. \textbf{6}, 2577 (1972); P. B. Allen and R. Silberglitt,
\emph{ibid}. \textbf{9}, 4733 (1974).

\bibitem{Yan2008} J. Yan, E. A. Henriksen, P. Kim, and A. Pinczuk,
Phys. Rev. Lett. \textbf{101}, 136804 (2008).

\bibitem{Park2008} C.-H. Park, F. Giustino, M. L. Cohen, and S. G. Louie, Phys. Rev. Lett. \textbf{99}, 086804 (2007); Nano Lett. (to
be published) (2008).

\bibitem{Rycerz2007} A. Rycerz, J. Tworzyd{\l}o, C. W. J. Beenakker, Nature Physics \textbf{3}, 172 (2007).

\bibitem{Akhmerov2007} A. R. Akhmerov and C. W. J. Beenakker, Phys. Rev. Lett. \textbf{98},
157003 (2007).

\bibitem{Woods2000} L. M. Woods, and G. D. Mahan, Phys. Rev. B \textbf{61},
10651 (2000).

\bibitem{Mahan2003} G. D. Mahan, Phys. Rev. B \textbf{68}, 125409 (2003).

\end{thebibliography}
\end{document}